\documentclass[preprint,5p,twocolumn]{elsarticle}

\def\RCSdate{February 29, 2018}
\date{\RCSdate}

\usepackage{amssymb}
\usepackage{amsmath}

\usepackage{booktabs} 
\usepackage[colorinlistoftodos]{todonotes}

\usepackage[ruled]{algorithm2e} 
\usepackage[utf8]{inputenc}

\usepackage{pgfplots}
\pgfplotsset{compat=1.14}
\usetikzlibrary{patterns}
\usepackage{subcaption}

\usepackage{makecell}
\usepackage{multirow}
\usepackage{colortbl}

\usepackage{xcolor}
\definecolor{delete}{rgb}{1, 0.0, 0}
\definecolor{domingo}{rgb}{.7, 0.4, 0}

\journal{Journal of Cultural Heritage}

\begin{document}

\begin{frontmatter}



\title{An Efficient GPU Approach for Designing 3D Cultural Heritage Information Systems}


\author[1]{Luis López\corref{cor1}}
\ead{luislopez@ugr.es}
\cortext[cor1]{Corresponding author}
\author[1]{Juan Carlos Torres}
\author[1]{Germán Arroyo}
\author[1]{Pedro Cano}
\author[1]{Domingo Martín}
\address[1]{Virtual Reality Laboratory. University of Granada. C/ Zamora, parcela 127-128. 18151 Ogíjares, Granada. Spain}

\begin{abstract}
We propose a new architecture for 3D information systems that takes advantage of the inherent parallelism of the GPUs. This new solution structures information as thematic layers, allowing a level of detail independent of the resolution of the meshes. Previous proposals of layer based systems present issues, both in terms of performance and storage, due to the use of octrees to index information. In contrast, our approach employs two-dimensional textures, highly efficient in GPU, to store and index information. In this article we describe this architecture and detail the GPU algorithms required to edit these layers. Finally, we present a performance comparison of our approach against an octree based system.
\end{abstract}


\end{frontmatter}

\section{Introduction}

Documentation of cultural heritage artefacts is one of the most important tasks in terms of understanding and preserving tangible heritage. This process usually involves handling huge sets of heterogeneous data: photographs, illustrations, recordings, logbooks, plans, diaries, databases, digitized 3D models, etc. In traditional information systems, these data are usually stored in the absence of any kind of cross-referencing between the 3D model and the database.\par   
The development of 3D scanning technologies over the last decade has allowed us to capture highly accurate representations of cultural heritage artefacts. Following the application of the proper processing techniques, the resulting 3D models possess sufficient geometric detail for them to be useful in terms of taking measurements and performing geometric operations.\par   
The geometric information offered by these digitized 3D models is connected with every other type of meaningful data used to document artefacts. After all, pictures are taken to accurately document specific areas, illustrations are drawn to emphasize finer details, logbooks are used to register the work done in different sections at specific times, etc. It thus seems natural, therefore, to organize and store this information on the surface of the 3D models.\par 

\section{Research Aim} \label{sec:research}

\begin{figure*}[!t]
\centering
\includegraphics[width=0.64\textwidth,keepaspectratio]{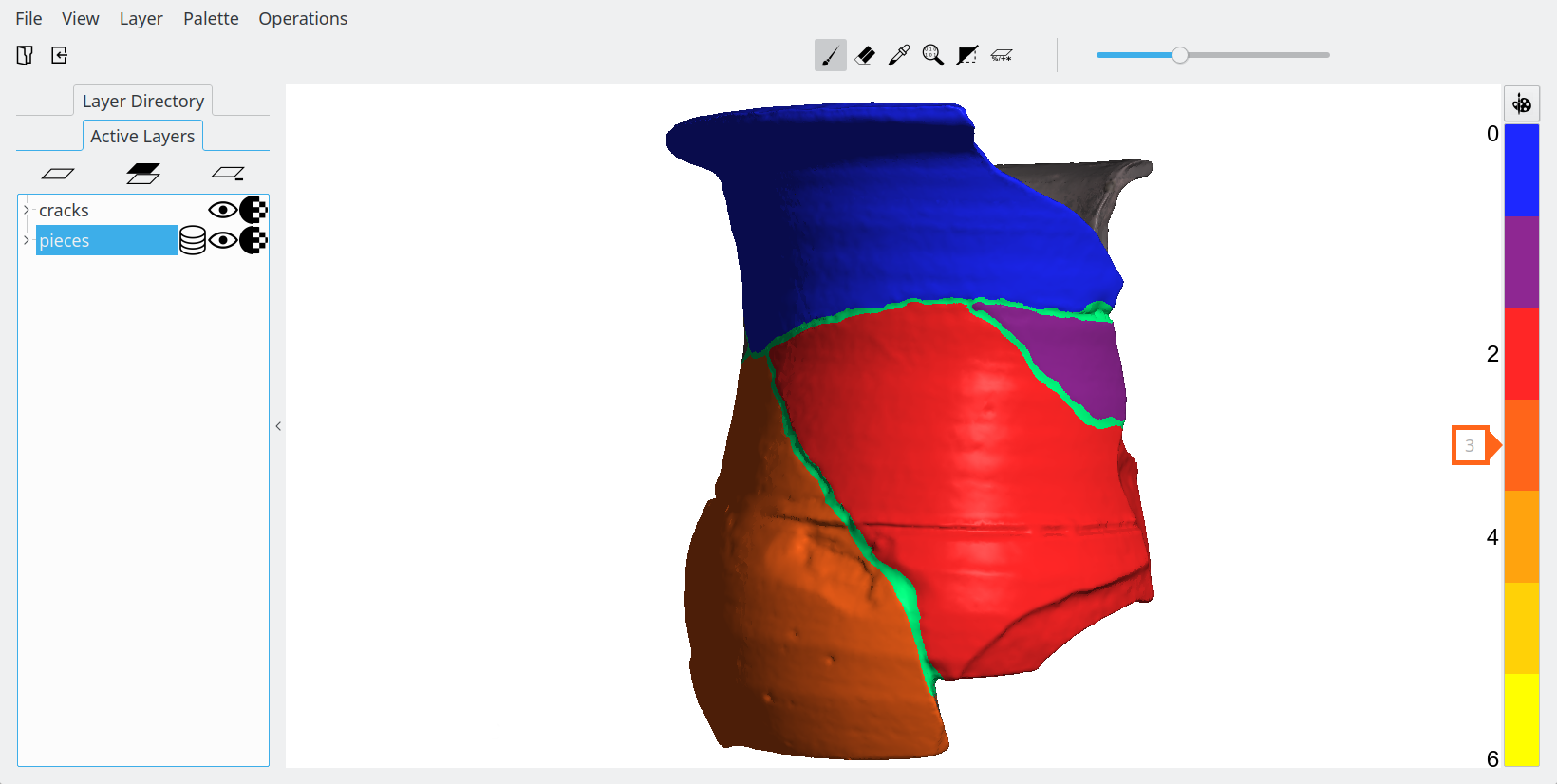}
\caption{Prototype of the proposed architecture. The user has associated two information layers to the 3D model of a vessel. The first layer \textit{cracks} identifies the area of the cracks using a green color. The second layer \textit{pieces} identifies the different pieces of the shattered vessel using the selected palette.}
\label{fig:app}
\end{figure*}

This article presents a new architecture for cultural heritage information systems that borrows from the foundations and functionality of Geographic Information Systems (GISs) and applies them to 3D models. Our solution organizes the information in thematic layers that are mapped onto the surface of the 3D model (Figure~\ref{fig:app}). The data of these layers is stored on 2D textures and texture coordinates are used to properly index that information. This new approach takes advantage of the parallelism and efficiency of the Graphics Processing Units (GPUs) to handle and operate these structures. Our approach also allows to associate information independently of the geometry of the 3D model.\par

\section{Previous works} \label{sec:prevworks}

In this section, we classify previous works related to information systems into four categories according to the dimension of the space employed in the analysis and visualization of data.

\subsection{2D space}

These methods represent the information by means of 2D structures and employ existing GISs with minor modifications to document tangible heritage. This choice provides a vast range of solid tools with powerful analytic and information retrieval capabilities. However, these methods lack a bidirectional connection between the stored information in a 2D plane and the digitized 3D source model. It is, therefore, necessary to process this 3D model before using it as a valid input in these systems.\par
Nagli{\v c}\cite{Naglic03} and Ioannidis et al.\cite{ioannidis2003} are good examples of this approach. They both use GISs for their work on large-scale archaeological sites, as these systems allow them to index large areas by geographical coordinates. Likewise, Parkinson et al.\cite{Parkinson2014} employ GISs to study tooth marking patterns created by large contemporary canids on the bones of their preys and compare them with earlier fossils. They photograph the bones and manually create vector layers of the markings. In each of these cases they have to compromise the 3D nature of the source material in order to use GISs to analyse the data, limiting their entire work-flow to just one point-of-view at a time.\par

\subsection{2.5D space}

These methods work with images and GISs to analyse and process the data. In contrast to the 2D approaches, the images are rasterized elevation models that contain height information. Therefore, they work in the restricted 3D space that this type of images provide. It is a more flexible and powerful method but, at the same time, it shares the issues of the 2D approaches. They need to convert the digitized 3D models into the appropriate image format and they are restricted to only one point of view.\par 
Benito et al.\cite{benito2015} use this approach to classify stone tools employed by wild chimpanzees. They divide this classification in several stages: first they scan the tools, then they transform the resulting 3d models into digital elevation models and finally, they use morphometric GIS classification functions to discriminate between active and passive pounders in lithic assemblages.\par

\subsection{Hybrid space}

The methods under this category work in a 2D space during the stages of analysis and data processing and then visualize the resulting information in a 3D space. Two different transformation steps are required for this process. First, they need to project the initial 3D model, used as geometric reference, onto different 2D planes. The resulting images or digital elevation models are later processed in a GIS. Once this process has finished, the output needs to be projected onto the 3D model again as a texture. Although these methods lack the point-of-view restrictions of those described above, their work-flow is more complex and it suited only to tangible heritage easily divisible in 2D planes.\par
Campanaro et al.\cite{Campanaro2016} used this approach to tackle the preservation of architectonic structures. They project the fa{\c{c}}ade of buildings into multiple images, process them in different GISs and finally, they project the results onto a simplified version of the original 3D models of the buildings for their visualization.\par

\subsection{3D space}

Unlike the previous approaches, these methods work directly with the digitized 3D models. There are no point-of-view restrictions and no transformations between different work spaces.
\subsubsection{Annotation systems}
The main goal of these systems is to associate information on specific sections of 3D model surface and offer robust information retrieval tools. There is a wide spectrum of indexation mechanisms under this paradigm, ranging from submeshes of the original 3D model to lines or points in 3D space.\par 
Durand et al.\cite{Durand2006}, Meyer et al.\cite{Meyer2007} and Mateos et al.\cite{Mateos2016} propose several online information systems to document archaeological sites. These systems require that the original 3D model is segmented into smaller and distinct entities. The analysis, processing and dissemination of information are done for each entity instead of the complete model.\par
Giunta et al.\cite{giunta2005} use AutoCAD models to structure the architectural information and diagnostic investigation results of Milan's Cathedral Fa{\c{c}}ade. Additionally, the system allows the user to insert pictures, texts and documents in a geo-referenced way.\par
Serna et al.\cite{serna2011} describe a distributed information system to annotate multimedia objects (3D models, images, text) using the concept of areas. Apollonio et al.\cite{Apollonio17} develop a web information system to document the restoration project of Neptune's Fountain in Bologne, Italy. The system allows the user to annotate the 3D model using three different primitives (points, polylines and areas) and to gather the stored data by means of robust information retrieval tools.\par
\subsubsection{Layer based systems}
These systems structure information attached to 3D models as a set of layers, where each layer stores the value of one attribute. Data layers can be considered as functions that associate a property value to points on the surface of the object. These type of systems can include the same functionality as the annotation systems and they can also perform complex operations between data layers. However, the main problem that these systems need to solve is how to design efficient structures and methods to represent those data layers.\par
Torres et al.\cite{Torres2010} divide the surface of the original 3D model into cells by means of an octree. Specifically, surfaces are recursively subdivided by eight cubic cells or voxels of the same size up to a predefined resolution level. Hence, each cell stores the triangles of the mesh that intersects. The level of detail depends on the size of these voxels and therefore, the number of subdivisions (levels) applied. The octree structure allows the user to work independently of the geometric irregularities and resolution. This way, fairly simple meshes are able to store information with better accuracy than the geometric mesh is able to offer. However, this spatial indexation becomes expensive in terms of memory and performance when dealing with the highest resolution levels.\par
The information layers are stored as sequences of properties assigned to the leaf nodes crossed by the surface\cite{Torres2013,Soler2013}. This system not only is able to annotate or look up information, but it also allows the user to make complex computations with heterogeneous layers. Some of
these computations include arithmetic, logic, geometric and topological operations or database queries, which can be used to analyze already existing data or to produce new information. Soler et al.\cite{Soler2017} improve this system in order to solve specific topological problems at the expense of using a more complex data structure.\par

In this article we propose a new solution for these layer based systems that works in the 3D space and uses information layers to organize the information annotated on the 3D model. Unlike the Torres et al.\cite{Torres2010} approach that requires an octree, our solution takes advantage of the modern GPU hardware by means of using textures and texture coordinates for data storage and indexation of the information. Next section describes in detail all the aspects of the data structures and algorithms required to create the proposed architecture.\par

\section{Proposed approach} \label{sec:approach}

This section offers a comprehensive look at our proposed architecture. Section~\ref{sec:textureHet} introduces the concept of using textures to store data other than colours. Section~\ref{sec:informationlayer} describes the layers and the rest of the structures. Finally, Section~\ref{sec:editinglayers} details the algorithms required to edit the layers.

\begin{figure}[!t]
\centering
\includegraphics[width=0.48\textwidth,keepaspectratio]{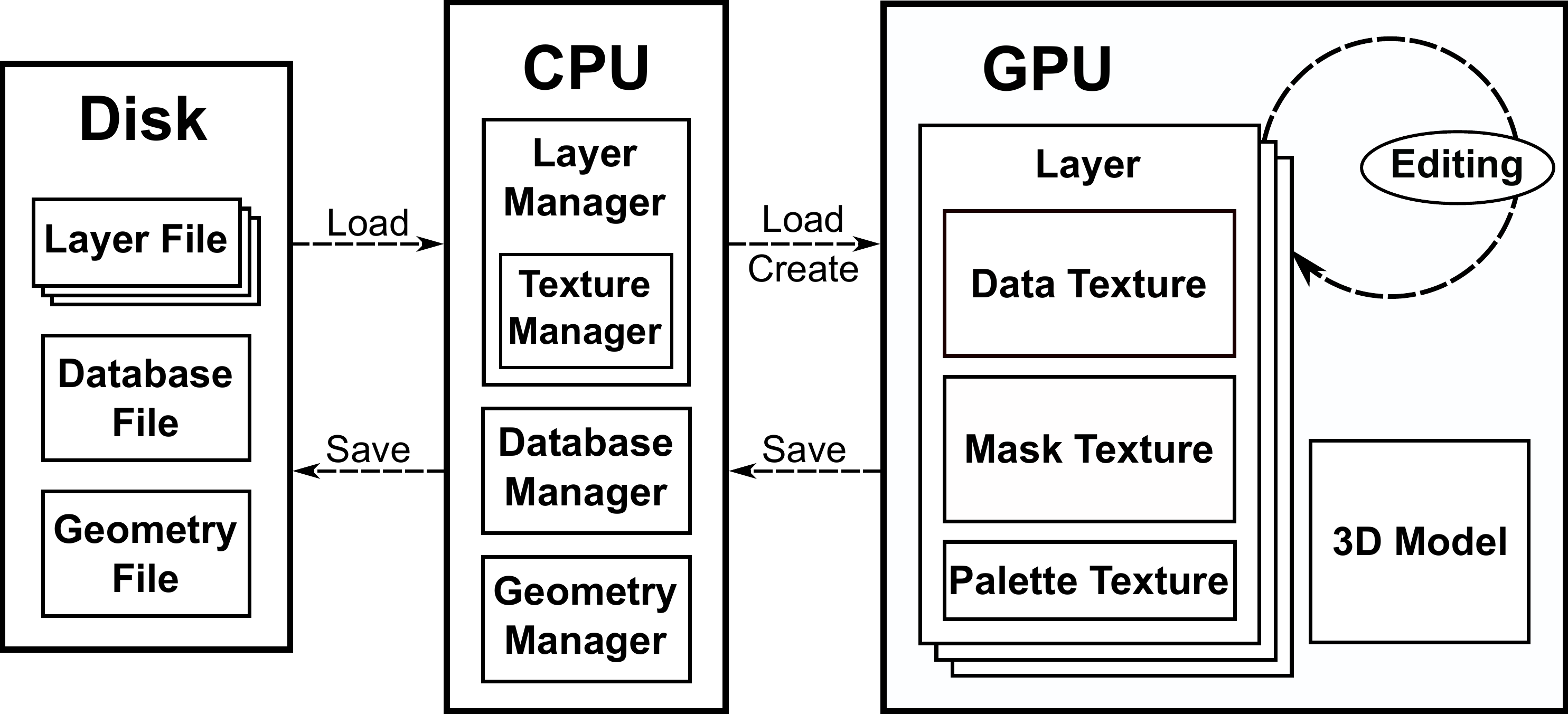}
\caption{General overview of the proposed architecture.}
\label{fig:general}
\end{figure}

We can organize our architecture around three computer components:
\begin{itemize}
\item \textbf{Disk drive} stores information permanently. It is the slowest component. 
\item \textbf{CPU} is a general purpose processor that solves computer tasks with the help of the primary memory. It is orders of magnitude faster than disk drives.
\item \textbf{GPU} is a specialized parallel processor that solves specific tasks. It is the fastest of these three types of components.
\end{itemize}

Ideally, whenever possible, structures and algorithms should be written with the strengths of the GPUs in mind to minimize the involvement of the CPU.\par

Our architecture (Figure~\ref{fig:general}) follows this principle and it defines all the structures directly in the memory of the GPU. Moreover, all the algorithms are solved in the GPU as well, so the CPU only has a management role. Layer Manager handles the loading, storage and creation of layers. Texture Manager is part of Layer Manager and issues commands to the GPU to create or destroy these structures but it does not store any of their information. Geometry Manager handles the loading and storage of geometry files and uploads the geometry data efficiently to the GPU. Database Manager resolves the creation and update of tables, tuples and queries. The disk drive is only used to load and store layers, databases and geometry permanently. Finally, the editing process requires a graphical user interface, therefore we have implemented a prototype (Figure~\ref{fig:app}) based in our architecture.\par

\subsection{Texture as a heterogeneous data structure} \label{sec:textureHet}

Textures are usually defined as containers of one or multiple images in computer graphics. These images are arrays of texels of a certain dimensionality (1D, 2D or 3D) and they store the information following a specific format. Traditionally, textures have been widely used to add color information to the surface of 3D models. However, the birth of the programmable graphics and GPU computing pipelines have dramatically changed their purpose during the last two decades.\par

Therefore, textures are multi-purpose structures nowadays. In fields like General Purpose GPU (GPGPU) computing, textures are usually treated as simple arrays or computation matrices. However, in Computer Graphics, it is a \textit{de facto} standard to work with 3D models and textures coordinates, which are necessary to access and store the information. These coordinates are the result of projecting every primitive of the mesh, such as triangles, onto the texture space, which is usually a 2D space.\par

Since our information system works with 3D models, we use texture coordinates to index the data stored in 2D textures. This data is represented as integers or floating numbers of a certain size based on the type of information layers.\par

\subsection{Information layer} \label{sec:informationlayer}

Our system handles two types of information layers:
\begin{itemize}
\item \textbf{Numeric layers} store numeric properties such as curvature, rugosity, age and so on. The properties can be integer or real values. These layers are normally employed for annotating and operating with quantitative information.
\item \textbf{Database layers} have a relational database table associated. Regardless of how the user defines the table schema, the primary keys are always unsigned integers and they are also the values stored in the layer. Therefore, these layers establish a bidirectional relationship between the table of a database and the geometry of the 3D model. The use of tables is an elegant solution that allows the storage of complex heterogeneous data (text, dates, documents, pictures) and greatly enhances the semantic information of the 3D models. These layers are used for annotating and operating with qualitative information.
\end{itemize}

Figure~\ref{fig:general} depicts how layers are implemented in our system. Specifically, information layers consist of two different 2D textures and one 1D texture:
\begin{itemize}
\item \textbf{Data texture} (2D) stores a number per texel. These numbers are integers (8, 16 or 32 bits) or float (16 or 32 bits) values. This texture holds the actual information of the layer, meaning property values in case of numeric layers, or primary keys in case of database layers. 
\item \textbf{Mask texture} (2D) stores a boolean value per texel to determinate whether the texel of the data texture contains valid information or not.
\item \textbf{Palette texture} (1D) stores the necessary color information per texel to visualize the contents of the data texture. Each texel holds a four component vector with the following color format: red, green, blue and alpha.
\end{itemize}
In order to manage the multiple data types and texture sizes, it is necessary to organize the textures efficiently because GPUs can only use simultaneously as many textures as the number of texture units they have. For instance, in modern NVIDIA graphics cards, this number is usually 32. Texture arrays are structures that can hold multiple textures and, at the same time, only occupy one texture unit. Therefore, they are extremely useful in cases like this. Our Texture Manager classifies the textures using a hash function and organizes them in texture arrays. The hash key is constructed by joining their width, height and data type in a unique field. Each new key creates a new texture array, whereas each texture that shares this key occupies a new layer of that texture array, avoiding any collision. A hash function is efficient due to the access operations are realised in constant order if there is no collisions. The textures always reside in GPU with the exception of storage and memory management operations. Texture Manager only stores the texture identifiers. Moreover, the process of editing the texture data is completely solved in GPU.\par
The commitment of these texture resources also happen dynamically in the GPU thanks to the use of sparse textures\cite{Everitt14}. This feature allows the separation of the GPU address space (reservation) from the requirement that all textures must be physically backed (commitment), exposing a limited form of virtualization for textures. Therefore, new resources can be reserved in the form of texture arrays with a huge number of layers without committing any GPU memory until it is required. The system does not require to reserve a specific number of each type of texture in advance, hence the flexibility of sparse textures allows more instances of different layer types simultaneously. Additionally, this technique alleviates the memory management and improves the performance due to there is no data-transfer between CPU and GPU until the physical memory of the GPU is almost fully occupied.\par
Our solution results in an efficient approach due to GPUs excel at working with textures and textures coordinates. Another advantage of this approach it is the independence between data and geometry in our model. Therefore, high resolution textures can be used to store information with a high level of detail on simple models.\par
Since these 2D textures do not store colour information but rather heterogeneous data, our system uses palettes that transform values into colours in order to render the layers. Our solution consists of a 1D texture that stores the colour information of the palette defined as a sequence of control points. Using the numeric value of the information layer per texel, another structure that contains the lower and upper limits of the palette and a single linear transformation, we can properly index the 1D texture and retrieve the correct colour to display.\par

\subsection{Editing layers on the 3D model} \label{sec:editinglayers}

Regarding to the user interface, once the editing mode is activated, an editing tool in the shape of a circle appears under the mouse cursor. The user can interactively add information to the selected layer by pressing the left button of the mouse and moving the tool over the desired area. This is a complex process that involves two distinct algorithms:
\begin{itemize}
\item Texture Editing Algorithm (TEA), detailed in Section~\ref{sec:textediting}, translates the user inputs into valid texture coordinates and it stores the new values in the appropriate texels of the layer textures.
\item Texture Padding Algorithm (TPA), described in Section~\ref{sec:textpadding}, eliminates the visual artefacts that could appear when the application renders the layer on the surface of the 3D model.
\end{itemize}

\subsubsection{Texture Editing Algorithm} \label{sec:textediting}

\begin{figure}[!htbp]
\centering
\includegraphics[height=0.91\textheight,keepaspectratio]{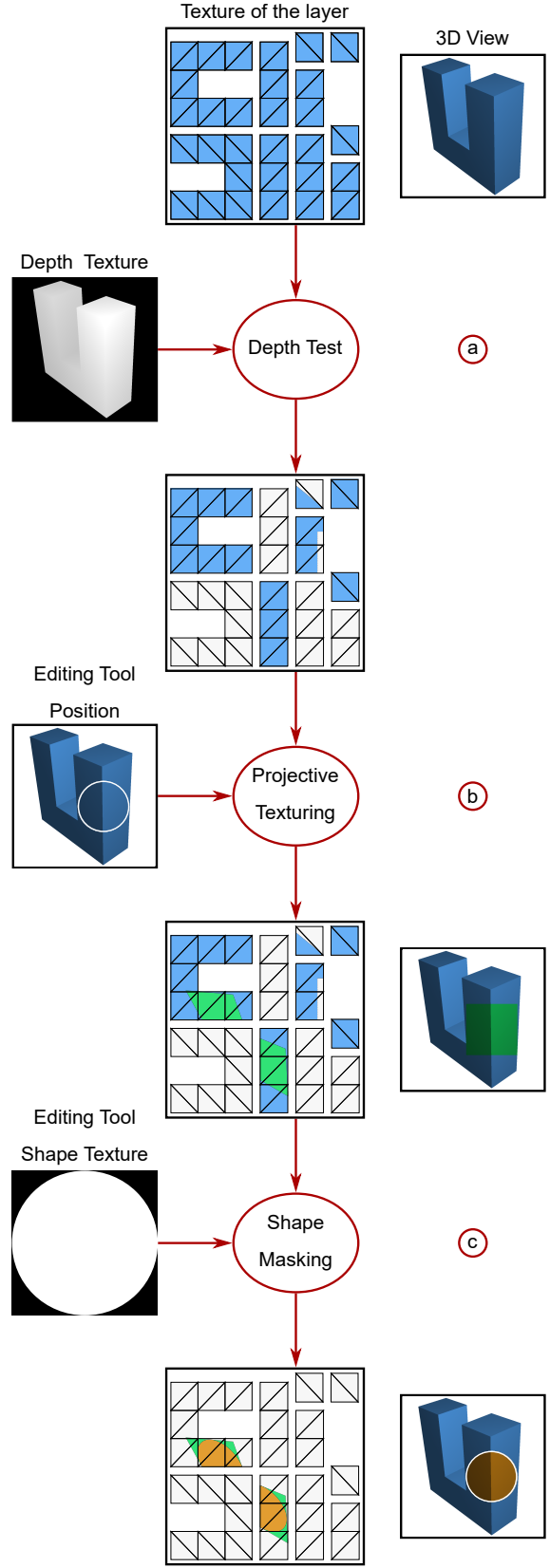}
\caption{Our algorithm filters areas of the texture progressively. (a) Depth Test discards the occluded parts of the 3D model. (b) Projective Texturing projects a squared area of the appropriated size onto the remaining area. Finally, (c) Shape Masking discards the texels that fall outside of the editing tool shape.}
\label{fig:texediting}
\end{figure}

TEA is a multi-step algorithm that projects the area of the editing tool, expressed in window coordinates, onto the space of the texture coordinates and stores the selected value on the appropriate textures permanently. Figure~\ref{fig:texediting} depicts a diagram of the whole process, which progressively discards regions of the texture until the edited area matches the orange shape of the editing tool.\par 

This algorithm requires two input textures in addition to other 3d model attributes, such as vertices, normals and texture coordinates and the information relative to the camera: 
\begin{itemize}
\item \textbf{Depth texture} is created as a part of the system pipeline. This texture stores the depth values of the scene from the current point of view. The user triggers an update when she modifies the viewpoint of the active camera or loads a new 3D model into the system.
\item \textbf{Editing tool shape texture} is created as a part of the system pipeline. This texture stores the shape of the current editing tool. The user triggers an update when she selects another shape for the editing tool.
\end{itemize}

In order to project the editing tool onto the 3D model surface at the position selected by the user, the algorithm makes use of projective texture mapping\cite{segal92}\cite{everitt01}. This technique assumes that textures are projected onto the scene by slide projectors. In this case, the projector and the scene camera share the same world position, hence its viewing transformation is also identical. Additionally, its projection transformation need to be adjusted to the texture frustum. If the projection transformation were the same, the texture would be projected onto the whole screen but that is not the goal of our algorithm. The projection has to be limited to the 2D area occupied by the editing tool. The basic equation for 2D coordinate transformation accomplishes that: 
\begin{equation*}
	T_{c} = T_{f} + S_{f} \cdot S_{c}
\end{equation*}
where $T_c$ is the target coordinate, $T_f$ is the translate factor, $S_f$ is the scale factor and $S_c$ is the source coordinate. The scaling and translate factors are computed with the following equations:
\begin{equation*}
  \begin{gathered}
  S_{f} = \left(\dfrac{W_{w}}{2T_{w}}, \dfrac{W_{h}}{2T_{h}}\right)\\
  T_{f} = \left(\dfrac{T_{px} - 0.5 \cdot W_{w}}{T_{w}}, \dfrac{T_{py} - 0.5 \cdot W_{h}}{T_{h}}\right)
  \end{gathered}
\end{equation*}
where $S_f$ and $T_f$ are the scale and translate factor respectively, $W_w$ and $W_h$ are the window size (width and height), $T_w$ and $T_h$ are the texture size (width and height), $T_{px}$ and $T_{py}$ are the horizontal and vertical coordinates of the editing tool.

Once the viewing and projection of the slide projector have been defined, the vertices of the 3D model are transformed to obtain the coordinates employed in the projective texture mapping. Unlike standard texture mapping that uses real texture coordinates $(s,t)$, projective texture mapping uses homogeneous coordinates or coordinates in the projective space $(s,t,w)$. These homogeneous coordinates are interpolated over the primitive and then projected onto the 2D texture space $(s/w, t/w)$ at each fragment before indexing the texture image. The domain for texture mapping is usually $[0,1]$ so those coordinates that fall outside of that range are discarded (Figure~\ref{fig:texediting}.b).\par 

At this point the accepted coordinates project a squared area onto the 3D model independently of the shape of the editing tool. Therefore, the coordinates need to be tested against the texture that stores the shape of the tool (Figure~\ref{fig:texediting}.c). This is a simple masking operation where the coordinates that fall outside the shape are also discarded. The final coordinates project the right shape onto the 3D model and the selected value is written on them.\par

The steps b) and c) project values onto the 3D model and visualize them using the active palette. However, the persistence of each editing operation is required and therefore the projected values need to be written on a texture and not rendered on the screen. In order to accomplish that, the default framebuffer is completely discarded. Instead, an off-screen framebuffer is set up and the 2D textures of the layer are attached to it. The scene is then rendered from the viewpoint of an orthogonal camera using the texture coordinates of the 3D model as vertex positions. That way, the texture coordinates are translated to the appropriate 2D texels in the texture space by means of perspective projection and the rasterisation process. Finally, only the texels that are validated by the different steps of the algorithm are updated: the selected value is stored on the data texture and the value \textbf{true} is stored on the mask texture of the current layer.\par

It is undesirable that the system allows the user to edit the occluded areas of the 3D models. This problem could appear because projective texture mapping does not carry out a depth test. To solve this issue, our algorithm makes use of the depth texture provided by the system pipeline. The texture stores the depth values of the scene viewed from the camera point-of-view. Before computing the projective texture mapping, each valid fragment is tested against the depth texture. A small bias is applied in this computation to fight possible precision issues. If the fragment falls behind the value of the pertinent texel, the fragment is discarded. Otherwise, the rest of the algorithm proceeds normally (Figure~\ref{fig:texediting}.a).\par 

\begin{figure}[!tbp]
\centering
\includegraphics[width = 0.48\textwidth]{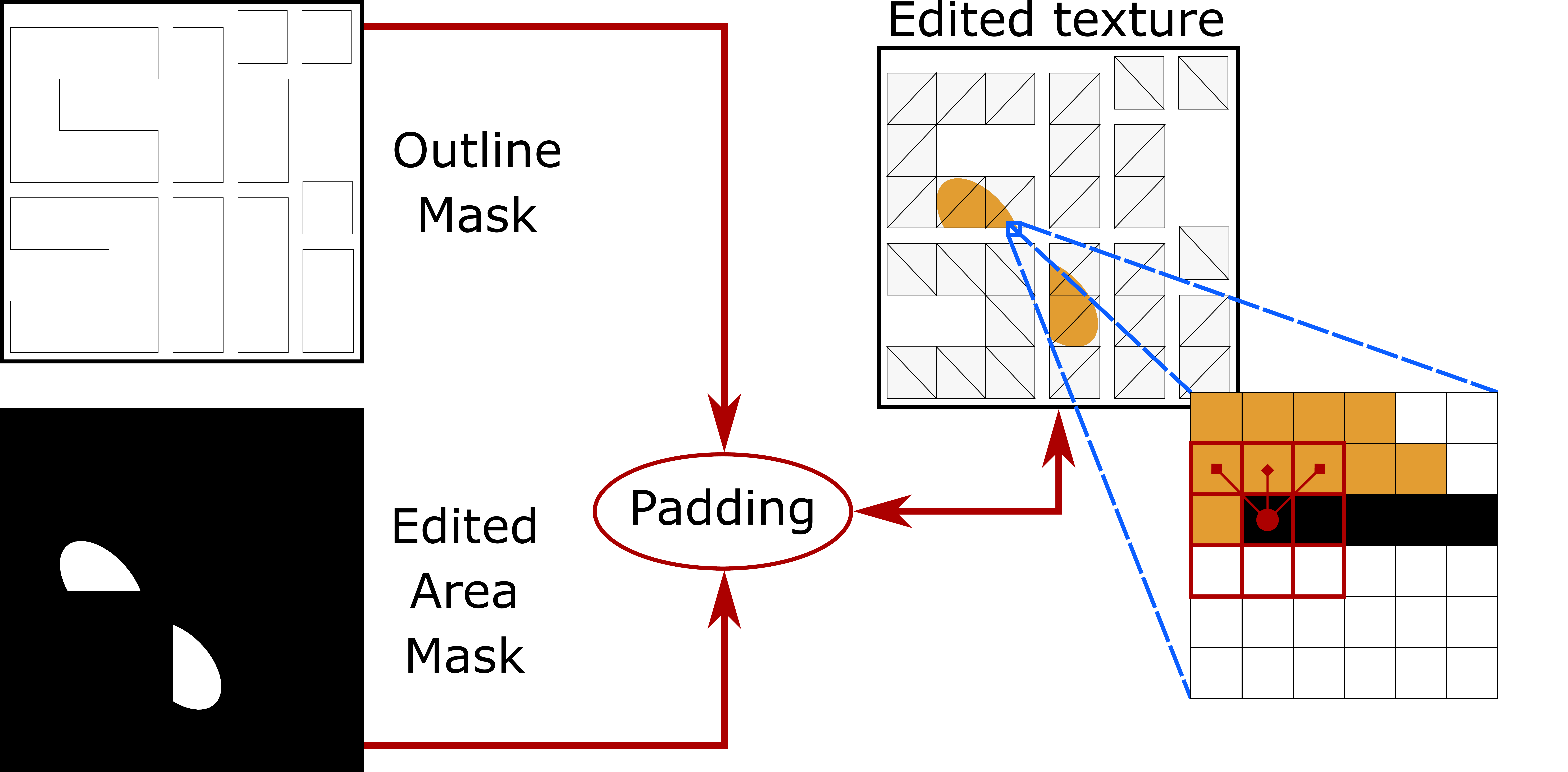}
\caption{In the zoomed area, the algorithm has verified that the current texel is part of the Outline Mask (black line) and the kernel (red matrix) is checking whether there are texels of the Edited Area Mask nearby (orange shape).}
\label{fig:texpadding}
\end{figure}

\subsubsection{Texture Padding Algorithm} \label{sec:textpadding}

While our TEA solves the editing of the data textures, the usage of textures also entails some visualization artefacts. When GPUs display textured 3D models, the texture coordinates define how the texels are sampled and they are usually organized as sets of isles of different size and shape along a 2D plane. The space of texture coordinates is continuous while the textures or images are discrete. This disparity makes the conversion between both spaces prone to small inaccuracies around the edge of the isles. If there is no redundant information beyond the borders of the texture isles, visual discontinuities or artefacts can appear when these borders are sampled. In order to prevent this rendering issue, our dilation algorithm expands the border data on some additional texels. The general idea of the process is shown in Figure~\ref{fig:texpadding}.\par

\begin{figure*}[!ht]
\begin{subfigure}{.24\textwidth}
  \centering
  \includegraphics[width=0.9\linewidth,height=.2\textheight, keepaspectratio]{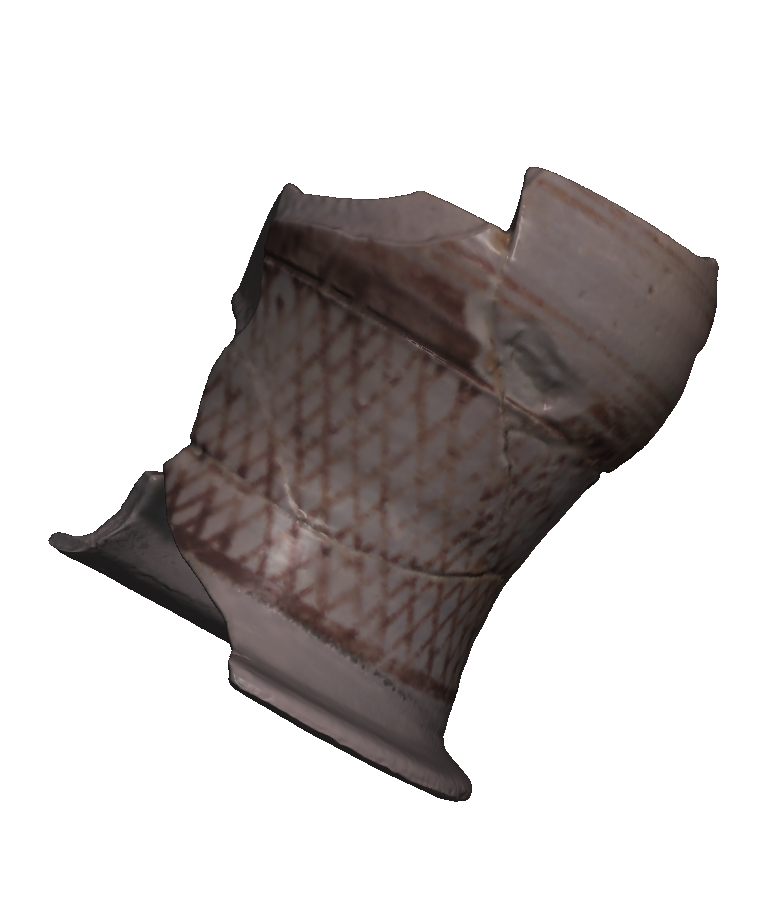}
  \captionsetup{justification=centering}
  \caption{Vessel\\312,094 triangles\\105.18 $\times$ 98.53 $\times$ 91.41 $mm$}
  \label{fig:mvessel}
\end{subfigure}%
\begin{subfigure}{.24\textwidth}
  \centering
  \includegraphics[height=.2\textheight, keepaspectratio]{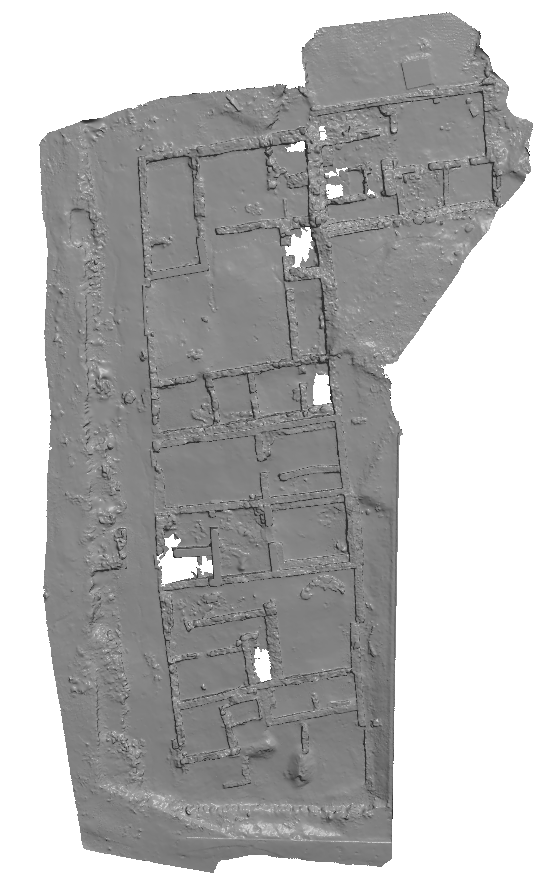}
  \captionsetup{justification=centering}
  \caption{Excavation\\999,942 triangles\\36.91 $\times$ 67.37 $\times$ 4.52 $m$}
  \label{fig:mpuente}
\end{subfigure}
\begin{subfigure}{.24\textwidth}
  \centering
  \includegraphics[height=.2\textheight, keepaspectratio]{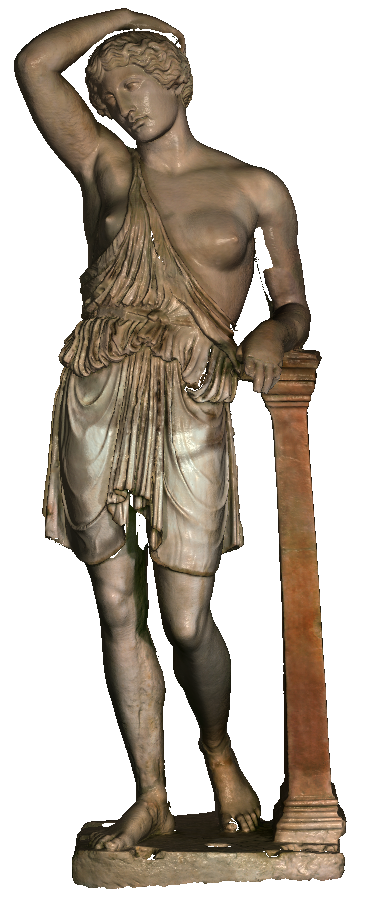}
  \captionsetup{justification=centering}
  \caption{Amazon\\5,711,204 triangles\\76.38 $\times$ 200.36 $\times$ 58.06 $mm$}
  \label{fig:mamazon}
\end{subfigure}
\begin{subfigure}{.24\textwidth}
  \centering
  \includegraphics[height=.2\textheight, keepaspectratio]{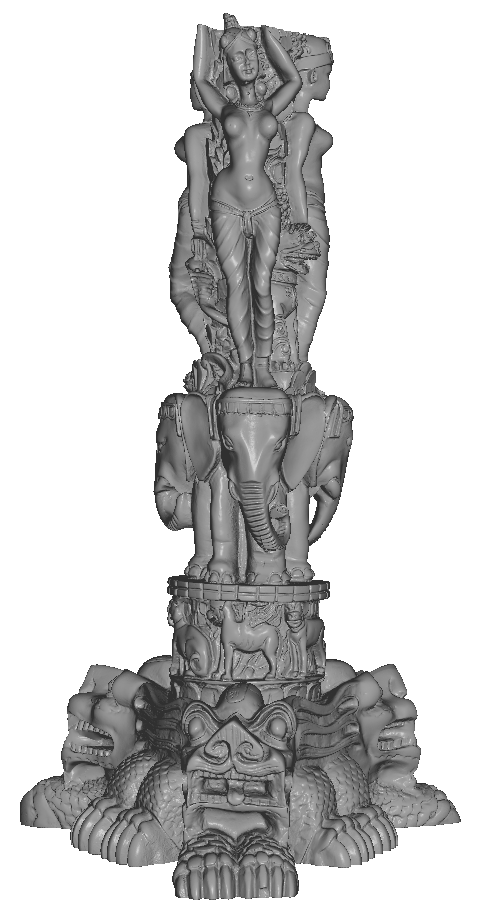}
  \captionsetup{justification=centering}
  \caption{Thai\\10,000,000 triangles\\235.22 $\times$ 396.04 $\times$ 203.16 $mm$}
  \label{fig:mthai}
\end{subfigure}
\caption{3D models used in the tests. Models (a) and (c) have colors per vertex while (b) and (d) use a generic gray color. They are organized in order of geometric complexity from left to right. The dimensions are also detailed for each model.}
\label{fig:3dmodels}
\end{figure*}

The algorithm requires two input textures:
\begin{itemize}
\item \textbf{Outline mask.} This texture contains the outlines of the texture islands. It is created as a part of the system pipeline by a two-pass algorithm. The first pass use the texture coordinates to render the texture parametrization of 3D model into a 2D texture. The second pass is a 2D image filter that use the output of the first pass and checks whether each texel is at a distance of the islands less than or equal to the intended thickness. It is updated every time a new 3D model is loaded into the system.
\item \textbf{Edited area mask.} This texture contains the shape of the edited area. It is attached to the same off-screen framebuffer used in TEA and therefore updated every time the user edits the layer.
\end{itemize}

This algorithm modifies the 2D textures of the layer edited in the previous subsection. The process is a quite straightforward image processing algorithm. It requires a kernel and the radius of the kernel is the width, in pixels, of the desired padding. Each texel of the texture is tested as follows: the algorithm checks whether the texel is part of the outline of a texture island. If that is the case, the algorithm checks then whether its distance to the edited area is less than or equal to the kernel radius. The texels that satisfy both conditions are part of the padding area and they are updated with the same value that the user selected to edit the layer.\par

This padding adds redundant information to the texture and guarantees the right colour when the GPU samples these conflicting areas. While this reduces the usable space in the texture, the issue is not significant enough to be too costly because our padding is only one texel wide.\par

\section{Results and discussion} \label{sec:results}

In this section we compare the performance of the prototype based on our architecture against another system based on an octree. Specifically, we selected the system designed by Torres et al.\cite{Torres2010}, noted as OCT-TR, because both systems work directly with 3D models, implement structures to associate information independently of the geometric complexity of the 3D models, use information layers to organize the data and therefore, they are very similar in terms of functionality.\par 

The hardware specifications of the computer used to conduct these tests are as follows: CPU Intel i7 4790k at 4.00 GHz, 16 GB DDR3 RAM memory at 1866 MHz and NVIDIA GTX 970 at 1.114 GHz with 4 GB GDDR5 RAM memory at 7 GHz.\par

All the tests measured the performance of the layer editing process in equivalent scenarios. For our solution, this involves the two algorithms explained in the last section: Texture Editing and Texture Padding. For OCT-TR, it involves the CPU casting multiples rays to find which voxels of the octree they collide with, updating the layer values accordingly and transferring the new version of the layer to the GPU.\par

We chose three levels of detail for the information layers to analyse the performance when the precision increases: three different texture resolutions for our solution and three different depths for OCT-TR. Moreover, for each one of those three cases, we selected five different editing tool sizes to analyse the performance when the edited area growths. All these tests used the same four 3D models (depicted in Figure~\ref{fig:3dmodels}).\par

\definecolor{customBlue}{HTML}{004282}
\definecolor{customRed}{HTML}{B01030}
\definecolor{customGreen}{HTML}{208320}

\pgfplotsset{
    legend image with text/.style={
        legend image code/.code={%
            \node[anchor=center] at (0.3cm,0cm) {#1};
        }
    },
}

\pgfplotscreateplotcyclelist{customcolor}
{
  	customGreen,thick,dashed,every mark/.append style={fill=customGreen}, mark=*, mark options={solid}\\
  	customRed,thick,dashed,every mark/.append style={fill=customRed},mark=*, mark options={solid}\\
  	customBlue,thick,dashed,every mark/.append style={fill=customBlue},mark=*, mark options={solid}\\  
	customGreen,thick,every mark/.append style={fill=customGreen}, mark=square*\\
  	customRed,thick,every mark/.append style={fill=customRed}, mark=square*\\
	customBlue,thick,every mark/.append style={fill=customBlue}, mark=square*\\    
}

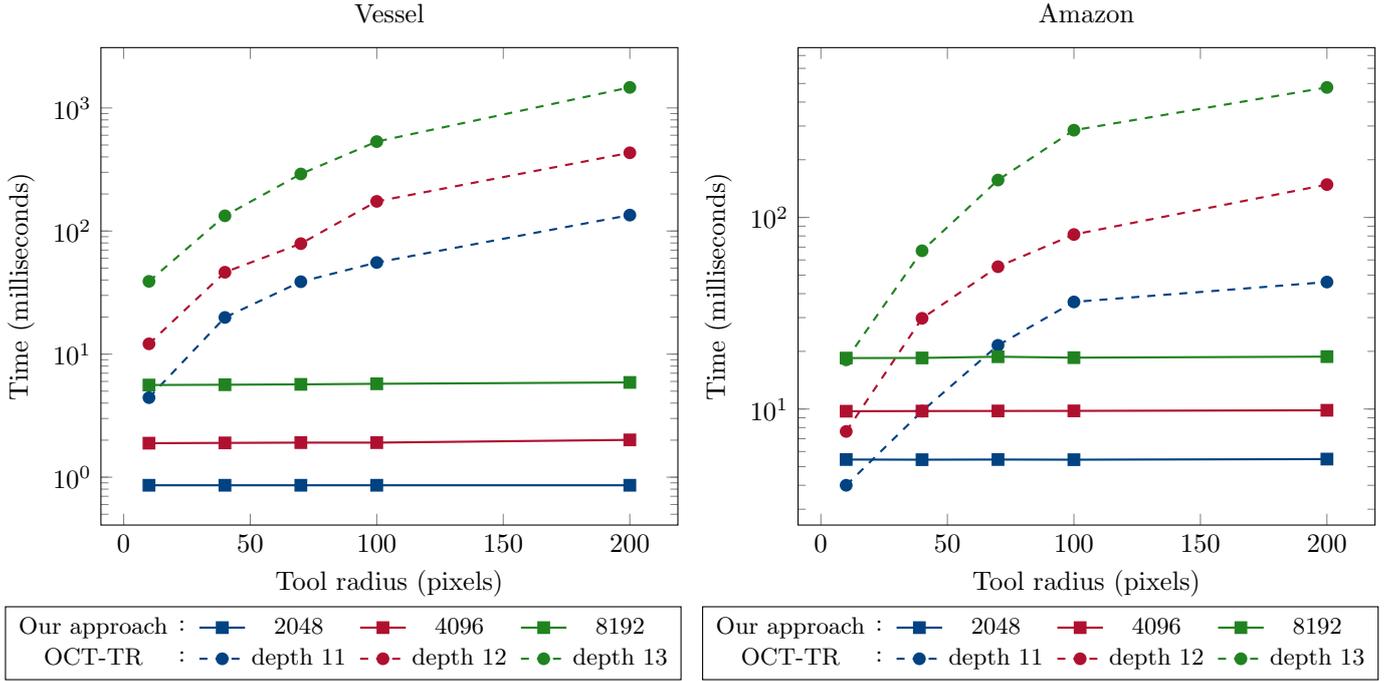
\begin{figure*}[!ht]
	\begin{subfigure}[t]{0.5\textwidth}
		\begin{tikzpicture}
			\begin{axis}
			[
				title=Vessel,
            	ymode=log,
            	width=1.0\textwidth,
      			legend style=
                {
                	font=\small,
                	reverse legend, at={(0.42,-0.17)}, anchor=north,legend columns=4, 									/tikz/every even column/.style={column sep=0.1cm}
                },
	  			xlabel={Tool radius (pixels)},        
      			ylabel={Time (milliseconds)},
                cycle list name=customcolor
    		]
			\addplot coordinates {(10,39) (40,133) (70,290.83) (100,533) (200,1469)};
            \addplot coordinates {(10,12.12) (40,46.25) (70,79) (100,174.14) (200,432.22)};
			\addplot coordinates {(10,4.43) (40,19.86) (70,38.75) (100,55.5) (200,134.75)};            
            \addlegendimage{legend image with text=OCT-TR}
            \addplot coordinates {(10,5.61) (40,5.64) (70,5.68) (100,5.74) (200,5.89)};
            \addplot coordinates {(10,1.89) (40,1.9) (70,1.91) (100,1.91) (200,2.01)};
            \addplot coordinates {(10,0.86) (40,0.86) (70,0.86) (100,0.86) (200,0.86)};
			\addlegendimage{legend image with text=Our approach}
            \legend{depth 13,depth 12,depth 11,:,8192,4096,2048,:}
			\end{axis}
		\end{tikzpicture}
		\label{fig:vessel}
	\end{subfigure}%
	\begin{subfigure}[t]{0.5\textwidth}
		\begin{tikzpicture}
            \begin{axis}
            [
            	title=Amazon,
            	ymode=log,
            	width=1.0\textwidth,                
               	legend style=
                {
                	font=\small,
                	reverse legend,at={(0.42,-0.17)},anchor=north,legend columns=4,
                    /tikz/every even column/.style={column sep=0.1cm}
                },
                xlabel={Tool radius (pixels)},        
            	ylabel={Time (milliseconds)},
                cycle list name=customcolor
            ]
            \addplot coordinates {(10,18.06) (40,67) (70,156.72) (100,285.28) (200,477)};            
            \addplot coordinates {(10,7.64) (40,29.75) (70,55.3) (100,81.43) (200,148.4)};
            \addplot coordinates {(10,4) (40,9.75) (70,21.5) (100,36.22) (200,46)};
            \addlegendimage{legend image with text=OCT-TR}
            \addplot coordinates {(10,18.45) (40,18.48) (70,18.75) (100,18.54) (200,18.77)};
            \addplot coordinates {(10,9.74) (40,9.76) (70,9.77) (100,9.78) (200,9.85)};
            \addplot coordinates {(10,5.45) (40,5.44) (70,5.45) (100,5.44) (200,5.48)};
            \addlegendimage{legend image with text=Our approach}
            \legend{depth 13,depth 12,depth 11,:,8192,4096,2048,:}
            \end{axis}
		\end{tikzpicture}
		\label{fig:amazon}
	\end{subfigure}%
    \caption{These graphs show the tests results, under logarithmic scale, for the editing of layers of two models: a) Vessel, b) Amazon. Tool radius corresponds to the radius of the editing tool in pixels. The solid lines correspond to our approach and the dashed lines to OCT-TR.}
    \label{fig:toolSizes}
\end{figure*}

\pgfplotsset{
  /pgfplots/bar legend/.style={
    /pgfplots/legend image code/.code={%
              \draw [#1] (0cm,-0.1cm) rectangle (0.2cm,0.1cm);},
  	/tikz/every even column/.style={column sep=0.2cm}              
}}

\pgfplotsset
{
  /pgfplots/bar cycle list/.style={/pgfplots/cycle list={
          	{customBlue,fill=customBlue,mark=none},
		  	{customBlue, fill=customBlue!20!white, postaction={pattern=north east lines,pattern color=customBlue},mark=none},
          	{customRed,fill=customRed,mark=none},
            {customRed, fill=customRed!20!white, postaction={pattern=north east lines,pattern color=customRed},mark=none},            
          	{customGreen,fill=customGreen,mark=none},
            {customGreen, fill=customGreen!20!white, postaction={pattern=north east lines,pattern color=customGreen},mark=none},
		},
	},
}

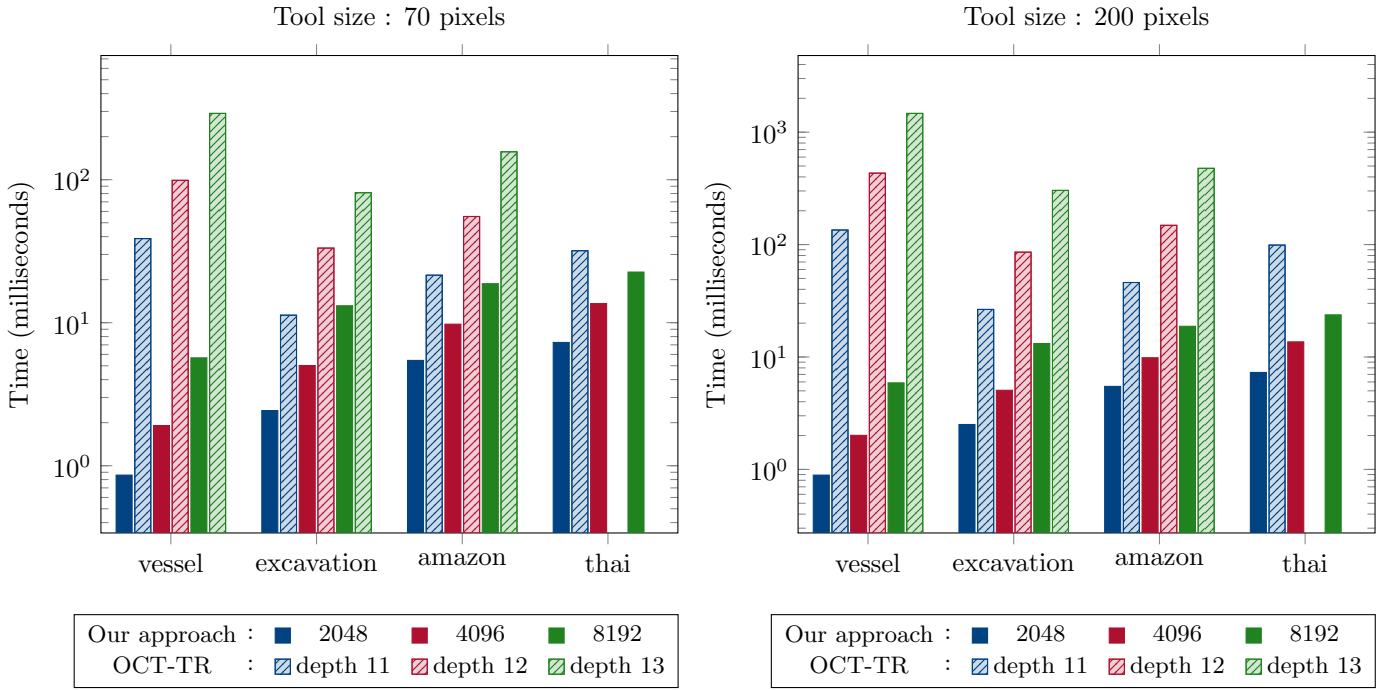
\begin{figure*}[!ht]
	\begin{subfigure}[t]{0.5\textwidth}
		\begin{tikzpicture}
			\begin{axis}
			[
            	title=Tool size : 70 pixels,
            	ymode=log,
                log origin=infty,
    			ybar=1pt,
                bar width=6pt,
                enlargelimits=0.16,
            	width=1.0\textwidth,
				legend columns=2, transpose legend,
      			legend style=
                {
                	font=\small,
                	at={(0.477,-0.17)}, anchor=north,legend columns=-1
				},
                bar legend,
	  			symbolic x coords={vessel,excavation,amazon,thai},       
      			ylabel={Time (milliseconds)},
                xtick=data                
    		]
            \addlegendimage{legend image with text=Our approach}
            \addlegendimage{legend image with text=OCT-TR}
			\addplot coordinates {(vessel,0.86) (excavation,2.43) (amazon,5.45) (thai,7.28)};            
            \addplot coordinates {(vessel,38.75) (excavation,11.29) (amazon,21.5) (thai,31.86)};
			\addplot coordinates {(vessel,1.91) (excavation,5.02) (amazon,9.77) (thai,13.61)};            
            \addplot coordinates {(vessel,99) (excavation,33.25) (amazon,55.3)};            
			\addplot coordinates {(vessel,5.68) (excavation,13.14) (amazon,18.75) (thai,22.61)};                        
            \addplot coordinates {(vessel,290.83) (excavation,81.11) (amazon,156.72)};
            \legend{:,:,2048,depth 11,4096,depth 12,8192,depth 13}
			\end{axis}
		\end{tikzpicture}
		\label{fig:modelsTex}
	\end{subfigure}%
	\begin{subfigure}[t]{0.5\textwidth}
		\begin{tikzpicture}
            \begin{axis}
			[
            	title=Tool size : 200 pixels,
            	ymode=log,
                log origin=infty,
    			ybar=1pt,
                bar width=6pt,
                enlargelimits=0.16,
            	width=1.0\textwidth,
				legend columns=2, transpose legend,
      			legend style=
                {
                	font=\small,
                	at={(0.477,-0.17)}, anchor=north,legend columns=-1
				},
                bar legend,
	  			symbolic x coords={vessel,excavation,amazon,thai},
      			ylabel={Time (milliseconds)},
                xtick=data,
    		]
            \addlegendimage{legend image with text=Our approach}
            \addlegendimage{legend image with text=OCT-TR}            
			\addplot coordinates {(vessel,0.89) (excavation,2.51) (amazon,5.48) (thai,7.29)};            
            \addplot coordinates {(vessel,134.75) (excavation,26.5) (amazon,46) (thai,99)};
			\addplot coordinates {(vessel,2.01) (excavation,5.06) (amazon,9.85) (thai,13.64)};            
            \addplot coordinates {(vessel,432.22) (excavation,85.66) (amazon,148.4)};            
			\addplot coordinates {(vessel,5.89) (excavation,13.18) (amazon,18.77) (thai,23.72)};                        
            \addplot coordinates {(vessel,1469) (excavation,303.42) (amazon,477)};            
            \legend{:,:,2048,depth 11,4096,depth 12,8192,depth 13}
            \end{axis}
		\end{tikzpicture}
		\label{fig:modelsOct}
	\end{subfigure}%
    \caption{These graphs show the tests results, under logarithmic scale, for the editing of layers using the the same tool size on each model. 70 pixel radius was used on the left graph and 200 pixel radius on the right graph. The 3D models are organized in order of geometric complexity. The solid colour bars correspond to our approach and the hatched bars to OCT-TR.}
    \label{fig:modelsPerf}
    
\end{figure*}

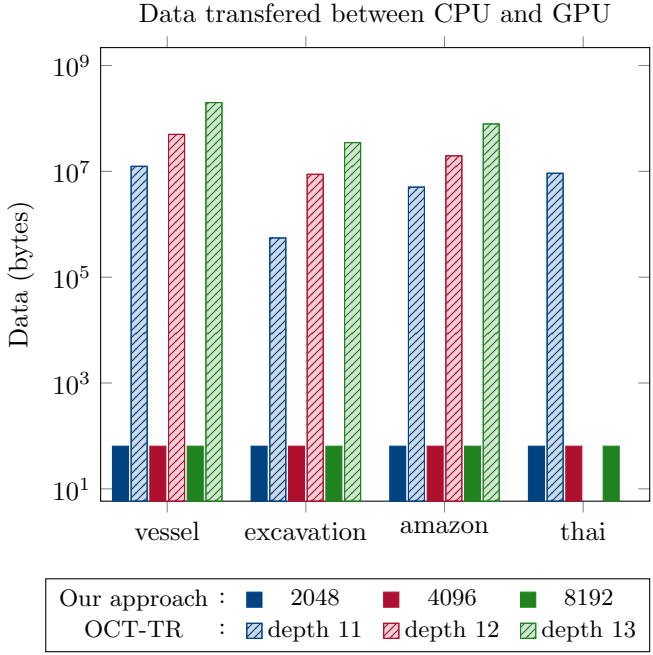
\begin{figure}
	\centering
	\begin{tikzpicture}
            \begin{axis}
			[
            	title=Data transfered between CPU and GPU,
            	ymode=log,
                log origin=infty,
    			ybar=1pt,
                bar width=6pt,
                enlargelimits=0.16,
            	width=0.48\textwidth,
				legend columns=2, transpose legend,
      			legend style=
                {
                	font=\small,
                	at={(0.450,-0.17)}, anchor=north,legend columns=-1
				},
                bar legend,
	  			symbolic x coords={vessel,excavation,amazon,thai},       
      			ylabel={Data (bytes)},
                xtick=data,
    		]
			\addlegendimage{legend image with text=Our approach}
            \addlegendimage{legend image with text=OCT-TR}
			\addplot coordinates {(vessel,64) (excavation,64) (amazon,64) (thai,64)};            
            \addplot coordinates {(vessel,12448544) (excavation,551368) (amazon,5038848) (thai,9199872)};
            \addplot coordinates {(vessel,64) (excavation,64) (amazon,64) (thai,64)};
            \addplot coordinates {(vessel,49948672) (excavation,8788000) (amazon,19652000)};                        
			\addplot coordinates {(vessel,64) (excavation,64) (amazon,64) (thai,64)};            
            \addplot coordinates {(vessel,199344128) (excavation,34967264) (amazon,78732000)};            	
            \legend{:,:,2048,depth 11,4096,depth 12,8192,depth 13}
            \end{axis}
	\end{tikzpicture}
    \caption{This graph show the amount of data, under logarithmic scale, that needs to be transferred between CPU and GPU during each editing operation. The solid colour bars correspond to our approach and the hatched bars to OCT-TR.}
    \label{fig:modelsTransf}
\end{figure}

In terms of level of detail, we performed multiple tests to establish the correspondence between texture resolutions and octree depths and empirically we reached to the following results: the 2048x2048 texture is equivalent to an octree depth of 11, the 4096x4096 texture to an octree depth of 12 and the 8192x8192 texture to the an octree depth of 13. Table~\ref{table:measurements} shows the precision, in square centimetres, guaranteed by both systems for the excavation model. The results for the other three models follow a similar pattern where our solution usually offers more precision than OCT-TR.

\begin{table}[!htbp]
\centering
\begin{tabular}{|l|c|c|c|}
\hline
\multirow{2}{*}{Method} & \cellcolor{gray!20} 2048 & \cellcolor{gray!20} 4096 & \cellcolor{gray!20} 8192 \\
						& depth 11 & depth 12 & depth 13 \\
\hline
    \rowcolor{gray!20} 	Our approach	& \multicolumn{1}{|r|}{24.71} &  \multicolumn{1}{|r|}{5.95} & \multicolumn{1}{|r|}{1.54} \\ \hline
                        OCT-TR  		& \multicolumn{1}{|r|}{81.15} & \multicolumn{1}{|r|}{23.45} & \multicolumn{1}{|r|}{4.86} \\
\hline
\end{tabular}
\caption{This table shows the precision guaranteed by both algorithms for the excavation model. The measurements are in square centimetres. The header of the columns shows the texture size (gray) and depth of octree (white) that are tested against each other.}
\label{table:measurements}
\end{table}

It is important to note that the tests of OCT-TR with the depths of 12 and 13 for the Thai statue could not be completed. This method required a high amount of memory to create the octree itself and the rest of its structures. The memory manager of the operative system showed that the system was using over 30 GB of virtual memory before the application crashed.

Figure~\ref{fig:toolSizes} shows how the performance evolves when the edited area changes under a logarithmic scale. After a detailed examination, we can conclude that our approach exhibits a significantly better behaviour: the results growth linearly in contrast with those of OCT-TR. Though the theoretical behaviour of our algorithm seems to be linear, this turn to be constant in practice due to the almost zero slope of the line, no matter the size of the editing tool. The reasons behind this excellent performance are explained by the good use of the GPU resources. Our algorithm allocates the workloads between GPU cores evenly and minimises the stalls in the GPU execution pipeline because there is no interdependencies in the calculations. All the operations are independent and inexpensive in terms of cost; they also use structures (textures) that are completely optimised by the architecture of GPUs. In contrast, the poor results obtained by OCT-TR are due to the own nature of the octree. Unlike our solution, this structure resides in the primary memory and the CPU computes all the operations. Concretely, the editing process involves the casting of multiple rays over the different voxels of the structure. The bigger is the editing area, the higher is the number of rays and collisions to check. The depth of the octree is also critical in terms of complexity because the voxels are halved by two in each dimension between consecutive depths and therefore, the number of collision tests growths exponentially.\par

Figure~\ref{fig:modelsPerf} shows, under a logarithmic scale, how the performance evolves when each 3D model is edited with an editing tool of the same size. After analysing the results, we conclude that our solution exhibits a better performance and a completely consistent behaviour across the four models. The time required to complete the editing operation increases when the 3D model becomes more complex or when the texture grows bigger. These results are reasonable because even though our solution takes advantage of the parallelism of GPUs, GPU resources are limited. At the same time, our algorithm performs better than expected: the difference between the results of the vessel and the Thai statue is always less than one order of magnitude even though there is a difference of almost two orders of magnitude in terms of geometry. In contrast, OCT-TR shows inconsistencies between models because the performance is highly dependent on how well balanced the octree is and which area is edited. When objects are projected onto octrees, one of their main features is the ability to discard complete octants in order to reduce the number of collisions to check. The shape of 3D models and changes in their orientation can make the central region of the octree heavily populated. Projecting the editing tool onto that region can make impossible to discard any octant in advance due to all of them contain sections of the model, affecting negatively to the performance. Therefore, the shape and orientation of 3D models are critical for spatial structures such as octress. The vessel and its sloped orientation is a perfect example of this disadvantage. It is the more demanding model even though it is the less complex in terms of geometry. Moreover, the performance is two orders of magnitude higher than our solution in the worse case tested, taking over a second to complete one single editing operation.\par

Figure~\ref{fig:modelsTransf} shows, under a logarithmic scale, the amount of data that our approach and OCT-TR transfer to the GPU for each editing operation. After analysing the results, it is evident that our approach performs significantly better. Since our layers always reside in GPU, our solution only transfer the 64 bytes of the matrix that represents the position of the editing tool. In contrast, OCT-TR uses two different representations for its layers: the primary memory stores the data and the GPU memory stores the colours. Therefore, OCT-TR have to send the updated version of the layer colours to the GPU in order to visualize the changes. These transfers require almost 200 Megabytes (MB) for the more detailed layers (depth 13) of the vessel every time an editing operation is performed.\par

Finally, our solution is usually more space efficient too. While our layers are bigger in size, our 3D model representation is more compact because it is constant in terms of space independently of the resolution of the layers. Therefore, our approach is better when the number of layers used simultaneously is below a threshold. Using the vessel as example, our system is more efficient in terms of space with less than eleven layers. Specifically, the size of our 3D model is 12 MB while OCT-TR requires 1.268 GB to store its model at depth of 13. However, for the highest resolution, the size of our layers is 327 MB while the size of OCT-TR layers is 199 MB.

\section {Conclusions} \label{sec:conclusions}

In this article we have proposed an efficient architecture for cultural heritage information systems. We also have carried out empirical tests comparing our approach to OCT-TR, clearly demonstrating that our representation is more efficient and can handle larger models. The strongest advantages of our approach are summarized in the following list:
\begin{itemize}
\item The size of our meshes is constant while OCT-TR subdivide its meshes when layer resolution is increased. This more compact format is valuable when researchers need to share information during field works.
\item The required time for the creation of our structures is insignificant in comparison with the creation time of the octree.
\item Our structures always reside in GPU.
\item All the operations are computed in GPU.
\item The data-transfers between CPU and GPU are closed to zero.
\item During the editing of the layers, the size of the editing tool does not affect the performance of the algorithm.
\end{itemize}

In conclusion, our architecture structures the information in thematic layers, uses 2D textures to store them and texture coordinates to index their information. Furthermore, it takes advantage of the inherent parallelism of GPUs to manage and operate these layers.

\section*{Acknowledgements}
This research was funded by the Spanish Ministry of Economy and Competitiveness (grants TIN2014-60965-R and TIN2017-85259-R) including FEDER funds from the European Union.

Thanks to Stanford Computer Graphics Laboratory, Museo de Puebla de Don Fadrique, Museo Histórico Municipal de Écija and Centro Andaluz de Arqueolog\'ia Ib\'erica for the 3D models used in this article.

\bibliographystyle{elsarticle-num}
\bibliography{bibliography}

\end{document}